%% file: main.tex
\documentclass[twocolumn]{aastex62}

\input{affil}

\received{2018 November 21}
\revised{2019 January 7}
\accepted{2019 January 7}
\published{2019 January 22}
\submitjournal{ApJL}

\shorttitle{Type~Ibn Supernovae May not all Come from Massive Stars}
\shortauthors{Hosseinzadeh et al.}

\begin{document}

\title{\bf \Large Type~Ibn Supernovae May not all Come from Massive Stars}

\correspondingauthor{Griffin Hosseinzadeh}
\email{griffin.hosseinzadeh@cfa.harvard.edu}

\author[0000-0002-0832-2974]{Griffin Hosseinzadeh}
\DSFP\CfA\LCO\UCSB

\author[0000-0001-5807-7893]{Curtis McCully}
\LCO

\author[0000-0001-6047-8469]{Ann I.~Zabludoff}
\UA

\author[0000-0001-7090-4898]{Iair Arcavi}
\TAU

\author[0000-0002-4235-7337]{K.~Decker French}
\Carnegie

\author[0000-0003-4253-656X]{D.~Andrew Howell}
\LCO\UCSB

\author[0000-0002-9392-9681]{Edo Berger}
\CfA

\author[0000-0002-1125-9187]{Daichi Hiramatsu}
\LCO\UCSB

\begin{abstract}
Because core-collapse supernovae are the explosions of massive stars, which have relatively short lifetimes, they occur almost exclusively in galaxies with active star formation. On the other hand, the Type~Ibn supernova PS1-12sk exploded in an environment that is much more typical of thermonuclear (Type~Ia) supernovae: on the outskirts of the brightest elliptical galaxy in a galaxy cluster. The lack of any obvious star formation at that location presented a challenge to models of Type~Ibn supernovae as the explosions of very massive Wolf--Rayet stars. Here we present a supplementary search for star formation at the site of PS1-12sk, now that the supernova has faded, via deep ultraviolet (UV) imaging of the host cluster with the \textit{Hubble Space Telescope}. We do not detect any UV emission within 1~kpc of the supernova location, which allows us deepen the limit on star formation rate by an order of magnitude compared to the original study on this event. In light of this new limit, we discuss whether or not the progenitors of Type~Ibn supernovae can be massive stars, and what reasonable alternatives have been proposed.

\end{abstract}


\section{Introduction} \label{sec:intro}
Core-collapse supernovae (SNe) are the terminal explosions of stars with zero-age main-sequence masses $M_\mathrm{ZAMS} \gtrsim 8\,\mathrm{M}_\odot$. Such stars have relatively short lifetimes: tens of millions of years for low-mass red supergiants down to a million years for high-mass Wolf--Rayet stars, assuming single-star evolution. Therefore, when we see these stars explode, they are almost always located in galaxies with recent or ongoing star formation. For example, in a sample of 2104 SN host galaxies, \cite{Hakobyan2012} found no core-collapse SNe in elliptical hosts, and only two in lenticular (S0) hosts.\footnote{Here we exclude the calcium-rich SNe~Ib, which may not be core-collapse SNe \citep{Perets2010}.} Both are SNe~II, the most common class of core-collapse SNe, and \cite{Hakobyan2012} suspected that both host galaxies have some residual star formation associated with diffuse or unresolved spiral arms.

A small fraction ($\lesssim$1\%) of SNe belong to a class whose spectra show little to no hydrogen and are dominated by narrow ($\sim$1000~km~s$^{-1}$) helium emission lines. \cite{Matheson2001}, \cite{Foley2007}, and \cite{Pastorello2007} first interpreted these spectral features as a sign of interaction between the SN ejecta and helium-rich circumstellar material (CSM). \cite{Pastorello2007} dubbed the class SNe~Ibn, in analogy to the much larger class of SNe~IIn, whose spectra indicate interaction with hydrogen-rich CSM. However, unlike SNe~IIn, whose light curves show a wide range of rise times and decline rates \citep{Kiewe2012}, SNe~Ibn have fast and relatively uniform light curves \citep{Hosseinzadeh2017c}, with some recent exceptions \citep{Pastorello2015b,Karamehmetoglu2017}. At the time of this writing, 31 SNe~Ibn have been classified.

Despite these differences, the preponderance of evidence has suggested that both SNe~IIn and SNe~Ibn come from very massive stars, whose strong winds eject a significant amount of material into the region around the star. When the star then explodes, the SN ejecta flash ionize and then collisionally excite the initially slow-moving CSM, leading to narrow emission lines in the SN spectra  (see \citealt{Smith2016} for a review). Hydrogen-rich SNe~IIn are thought to come from hydrogen-rich luminous blue variable stars, while hydrogen-poor SNe~Ibn have been attributed to hydrogen-poor Wolf--Rayet stars. While the former connection has been established by direct progenitor detection (\citealt{Gal-Yam2009,Smith2010a,Smith2010,Foley2011}; but see \citealt{Fox2017}), the only direct evidence for the latter comes from SN~2006jc, which was preceded by an outburst at the same location in 2004 \citep{Pastorello2007}.

\cite{Sanders2013} first reported on the mystery of PS1-12sk, the only SN~Ibn not to occur in a star-forming galaxy. Instead, the SN exploded in the galaxy cluster RXC~J0844.9+4258 at matching redshift \citep[$z=0.054$; ][]{Bohringer2000}, with the most likely host being the elliptical brightest cluster galaxy (BCG) CGCG~208--042, at a projected separation of 28~kpc. They detected no source at the SN location to a limit of $M_r \gtrsim -10.5$ and no narrow H$\alpha$ emission (from a potential host galaxy) in a spectrum of the SN to a 3$\sigma$ limit of $L_\mathrm{H\alpha} \lesssim 2 \times 10^{38}\mathrm{\,erg\,s^{-1}\,kpc^{-2}}$. The latter corresponds to a limit on the star formation rate density of $2 \times 10^{-3}\mathrm{\,M_\odot\,yr^{-1}\,kpc^{-2}}$, which is low but not unheard of for the site of a core-collapse SN \citep[see Figure~\ref{fig:limits};][]{Kelly2012,Galbany2018}. Nonetheless, \cite{Sanders2013} explored scenarios in which white dwarfs, which do not require recent star formation, could explode as SNe~Ibn. Photometry and spectra of PS1-12sk match quite well with other members of its class \citep{Hosseinzadeh2017c}; only its host environment stands out.

Here we report deep ultraviolet (UV) imaging of the host cluster of PS1-12sk from the \textit{Hubble Space Telescope} (\textit{HST}) taken six years after the SN has faded. The continued nondetection of a source at the SN location allows us to strengthen the limit on star formation by an order of magnitude, moving it below all previous core-collapse SNe. This prompts a serious consideration of the possibility that PS1-12sk, and by association other SNe~Ibn, do not come from massive stars.

\begin{figure*}
    \centering
    \includegraphics[width=0.95\textwidth]{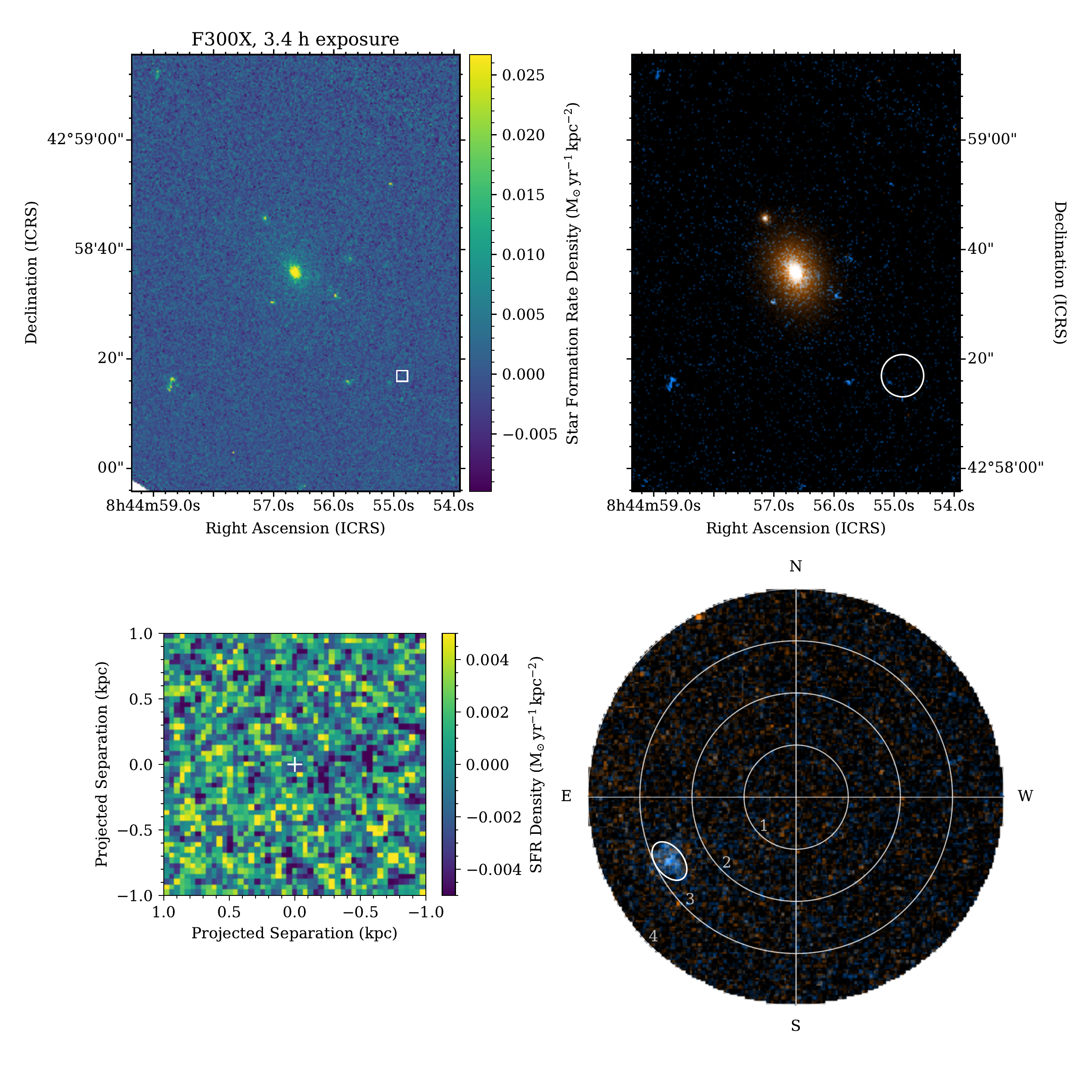}
    \caption{Top left: \textit{HST} images of the host cluster of PS1-12sk in F300X. The white square is a $2\,\mathrm{kpc} \times 2\,\mathrm{kpc}$ aperture around the SN location, inside of which we measure the star formation rate density.
    Bottom left: a $2\,\mathrm{kpc} \times 2\,\mathrm{kpc}$ region around the SN location (marked with a +) in F300X, i.e., the contents of the white square in the top left panel, enlarged and rescaled to show the lack of any sources.
    Top right: arcsinh-scaled pseudocolor image of the host cluster of PS1-12sk, where red represents F625W, blue represents F300X, and green is the average of the two filters. The white circle is centered on the SN location and has a radius of 4~kpc at the cluster redshift.
    Bottom right: a 4~kpc radius region around the SN location, i.e., the contents of the white circle in the top right panel. A white ellipse marks the nearest source, likely an ultra-compact dwarf galaxy if it is associated with the cluster.}
    \label{fig:images}
\end{figure*}

\section{Observations and Data Reduction} \label{sec:obs}
We obtained images, logged in Table~\ref{tab:obs}, with the UVIS channel of \textit{HST}'s Wide Field Camera~3 \citep{Dressel2018} under program GO-15236 (PI: G.~\citeauthor{2017hst..prop15236H}): a total exposure time of 3.4~hours with F300X, an extremely wide UV filter ($280.7 \pm 33.2$~nm), and 17.5~minutes with F625W, similar to the Sloan Digital Sky Survey (SDSS) $r'$ filter ($624.2 \pm 73.2$~nm). We downloaded the calibrated, charge transfer efficiency corrected \citep[\texttt{*\_flc.fits};][]{Gennaro2018} files from the Barbara A.~Mikulski Archive for Space Telescopes.

\begin{deluxetable}{cccc}
\tablecaption{Observation Log\label{tab:obs}}
\tablehead{\colhead{MJD} & \colhead{Filter} & \colhead{Exposure} & \colhead{Phase\tablenotemark{a}} \\[-6pt]
\colhead{at Start} & \colhead{} & \colhead{Time (s)} & \colhead{(days)}}
\startdata
58198.21573 & F300X & 896 & +2079.8 \\
58198.22765 & F300X & 896 & +2079.8 \\
58198.28063 & F300X & 896 & +2079.9 \\
58198.54669 & F300X & 896 & +2080.1 \\
58198.55861 & F300X & 896 & +2080.1 \\
58198.57053 & F300X & 896 & +2080.1 \\
58200.46627 & F300X & 896 & +2081.9 \\
58200.47819 & F300X & 896 & +2082.0 \\
58200.49011 & F300X & 896 & +2082.0 \\
58255.22808 & F300X & 896 & +2133.9 \\
58255.24000 & F300X & 896 & +2133.9 \\
58255.25192 & F300X & 896 & +2133.9 \\
58255.75761 & F300X & 675 & +2134.4 \\
58255.76698 & F300X & 675 & +2134.4 \\
58255.77664 & F625W & 350 & +2134.4 \\
58255.78221 & F625W & 350 & +2134.4 \\
58255.78778 & F625W & 350 & +2134.4 \\
\enddata
\tablenotetext{a}{Time after the $z$-band peak of PS1-12sk \citep[MJD 56006.1;][]{Sanders2013} in the cluster rest frame.}
\end{deluxetable}

We then combined the images for each filter using SNHST \citep{McCully2018}, an open-source pipeline for stacking \textit{HST} images. SNHST begins by refining the astrometric solution for a single reference visit to match a ground-based catalog; in this case, we matched our F625W images to the Panoramic Survey Telescope And Rapid Response System 1 \citep[Pan-STARRS1;][]{Chambers2016} $3\pi$ Survey catalog. It then refines the astrometric solutions for the other visits to match the reference visit. Once all the images have consistent astrometry, SNHST aligns and coadds them by filter, and optionally by visit, using DrizzlePac \citep{Gonzaga2012}. Finally it removes cosmic rays from the final images using Astro-SCRAPPY \citep{McCully2014}, which is based on the L.A. Cosmic algorithm \citep{vanDokkum2001}. Our final stacked images are shown at the top of Figure~\ref{fig:images}.

\defcitealias{Kennicutt1998}{Kennicutt's (1998)}

To be consistent with \cite{Sanders2013}, we assume a standard $\Lambda$CDM cosmology with $H_0 = 71\,\mathrm{km\,s^{-1}\,Mpc^{-1}}$, $\Omega_\Lambda = 0.7$, and $\Omega_\mathrm{m} = 0.3$, which yields a luminosity distance of 238~Mpc for the host cluster of PS1-12sk. We correct our measurements using the \cite{Schlafly2011} Milky Way extinction coefficients in the direction of PS1-12sk: $A_\mathrm{F300X} = 0.162$~mag and $A_\mathrm{F625W} = 0.070$~mag. Observations of PS1-12sk suggest that host extinction is negligible \citep[and Appendix~\ref{sec:extinct}]{Sanders2013}. To convert luminosity to star formation rate, we use \citetalias{Kennicutt1998} rescaling of the relationship from \cite{Madau1998} for a \cite{Salpeter1955} initial mass function integrated from 0.1 to 100~M$_\odot$:
\begin{equation}
\frac{\textrm{SFR}}{\mathrm{M_\sun\,yr^{-1}}} = \frac{L_\nu}{7.1 \times 10^{20}\,\mathrm{W\,Hz^{-1}}},
\label{eq:sfr}
\end{equation}
where $L_\nu$ is the average luminosity spectral density in a rectangular filter centered at 280~nm with a bandwidth of 56~nm. At the redshift of PS1-12sk, F300X has a pivot wavelength \citep{Koornneef1986} of 295.9~nm and a rectangular bandwidth of 69.9~nm, similar enough that we do not use a $K$-correction. Dividing both sides of Equation~(\ref{eq:sfr}) by area and rewriting luminosity in terms of surface brightness, we arrive at
\begin{equation}
-2.5\log_{10}\left(\frac{\Sigma_\mathrm{SFR}}{\mathrm{M_\odot\,yr^{-1}\,kpc^{-2}}}\right) = \Sigma_\mathrm{UV} - 18.5,
\label{eq:sfrd}
\end{equation}
\begin{equation}
\mathrm{where}\quad \Sigma_\mathrm{UV} \equiv \sigma_\mathrm{UV} - 10\log_{10}(1+z) - A_\mathrm{UV}
\end{equation}
is the redshift- and extinction-corrected UV surface brightness in AB magnitudes. (AB magnitudes are used throughout this Letter.)

There is no source visible within 1~kpc of the location of PS1-12sk in the final stacked F300X image. To quantify this, we measure the background-subtracted (see Appendix~\ref{sec:bkg}) flux in a $2\,\mathrm{kpc} \times 2\,\mathrm{kpc}$ square around the location of PS1-12sk (Figure~\ref{fig:images}, bottom left), which is consistent with zero. Using the Poisson uncertainty on the total number of electrons in the aperture, we calculate a $5\sigma$ limiting surface brightness of 27.5~mag~arcsec$^{-2}$, corresponding to a star formation rate density below $3.6 \times 10^{-4}\,\mathrm{M_\odot\,yr^{-1}\,kpc^{-2}}$. We also run SEP \citep{Barbary2016}, a Python implementation of Source Extractor \citep{Bertin1996}, on this portion of the image and find no sources at $3\sigma$ significance. Smoothing the image with a Gaussian kernel with $\sigma$ between 1 and 5 pixels does not change the result or reveal an obvious surface brightness gradient.

We also investigate the nature of the source closest to the SN location, which \cite{Sanders2013} suggested is likely an ultra-compact dwarf galaxy in the cluster, despite its unknown redshift. With \textit{HST}'s resolution, we refine the position of this source to be 2.7~kpc from the SN location, slightly further than the 2.4~kpc reported by \cite{Sanders2013}. The source is clearly detected in our F300X image and marginally detected in our F625W image, but is only marginally resolved even by \textit{HST} (Figure~\ref{fig:images}, bottom right). Using SEP, we measure a \cite{Kron1980} radius of about 120~pc, if the source is associated with the galaxy cluster, and Kron magnitudes of $\mathrm{F300X} = 24.78 \pm 0.05\,\mathrm{mag}$ ($M_\mathrm{F300X} = -12.31\,\mathrm{mag}$) and $\mathrm{F625W} = 24.3 \pm 0.1\,\mathrm{mag}$ ($M_\mathrm{F625W} = -12.6\,\mathrm{mag}$), the latter of which agrees with the measurement from the Pan-STARRS1 image: $r = 24.3 \pm 0.1\,\mathrm{mag}$ \citep{Sanders2013}. Based on these measurements and those of \citet[see their Figure~7]{Sanders2013}, we conclude that the source is indeed part of the cluster.

\section{Host Galaxies of Other SNe~Ibn}
To understand the context for our limit on the star formation rate near PS1-12sk, we examine the host galaxies of other SNe~Ibn in the literature. Here and throughout this Letter, we include transitional SNe~Ibn/IIn \citep{Pastorello2008}, whose spectra show some hydrogen lines but are still dominated by helium. Previously, \cite{Pastorello2015a} listed the morphologies, $B$-magnitudes, normalized offsets, and metallicities of a sample of 16 SN~Ibn host galaxies, and \cite{Taddia2015} measured the host galaxy metallicities for a sample of interacting SNe, including 6 SNe~Ibn. Other than the fact that they all occurred in spiral galaxies (except PS1-12sk), no obvious patterns emerge. As far as we are aware, no previous authors have systematically studied the star formation rates at the locations of SNe~Ibn.

\begin{deluxetable*}{lccCcl}
\tablecaption{Surface Brightnesses of SN~Ibn Hosts in SDSS Fields\label{tab:hosts}}
\tablehead{\colhead{Name} & \colhead{Redshift} & \colhead{Extinction\tablenotemark{a}} & \colhead{Surface Brightness} & \colhead{Aperture} & \colhead{Discovery/Classification} \\[-6pt]
\colhead{} & \colhead{$z$} & \colhead{$A_{u'}$ (mag)} & \colhead{$\sigma_{u'}$ (mag arcsec$^{-2}$)} & \colhead{(kpc)} & \colhead{Reference(s)}}
\startdata
SN 2002ao & 0.005 & 0.185 & 23.54 \pm 0.08 & 0.20 & \cite{Martin2002} \\
SN 2006jc & 0.006 & 0.085 & 24.05 \pm 0.25 & 0.24 & \cite{Nakano2006} \\
SN 2010al & 0.017 & 0.199 & 22.38 \pm 0.04 & 0.68 & \cite{Rich2010} \\
PTF11rfh & 0.060 & 0.220 & 24.20 \pm 0.12 & 2.26 & \cite{Hosseinzadeh2017c} \\
PS1-12sk & 0.054 & 0.129 & >24.28 & 2.05 & \cite{Sanders2013} \\
LSQ12btw & 0.058 & 0.096 & 22.58 \pm 0.08 & 2.19 & \cite{Valenti2012} \\
PTF12ldy & 0.106 & 0.250 & >24.15 & 3.79 & \cite{Hosseinzadeh2017c} \\
iPTF13beo & 0.091 & 0.185 & 23.77 \pm 0.12 & 3.31 & \cite{Gorbikov2014} \\
iPTF14aki & 0.064 & 0.137 & >24.98 & 2.40 & \cite{Polshaw2014,Hosseinzadeh2017c} \\
SN 2014av & 0.030 & 0.073 & 23.24 \pm 0.07 & 1.17 & \cite{Ciabattari2014} \\
SN 2014bk & 0.070 & 0.092 & 22.72 \pm 0.09 & 2.61 & \cite{Morokuma2014} \\
ASASSN-14ms & 0.054 & 0.049 & >24.77 & 2.05 & \cite{Kiyota2014,Vallely2018} \\
SN 2015U & 0.014 & 0.246 & 22.55 \pm 0.04 & 0.56 & \cite{Kumar2015,Ochner2015} \\
ASASSN-15ed & 0.049 & 0.104 & 23.54 \pm 0.11 & 1.87 & \cite{Fernandez2015,Noebauer2015} \\
iPTF15ul & 0.066 & 0.087 & 21.51 \pm 0.03 & 2.47 & \cite{Hosseinzadeh2017c} \\
iPTF15akq & 0.109 & 0.062 & >24.80 & 3.89 & \cite{Hosseinzadeh2017c} \\
SN 2016Q & 0.103 & 0.156 & 22.96 \pm 0.08 & 3.70 & \cite{Hounsell2016,2016TNSTR..32....1W,2016TNSCR..41....1P} \\
\enddata
\tablenotetext{a}{\cite{Schlafly2011}}
\end{deluxetable*}

\defcitealias{SDSSCollaboration2017}{SDSS Collaboration 2017}

\begin{figure*}
    \centering
    \includegraphics[width=\textwidth]{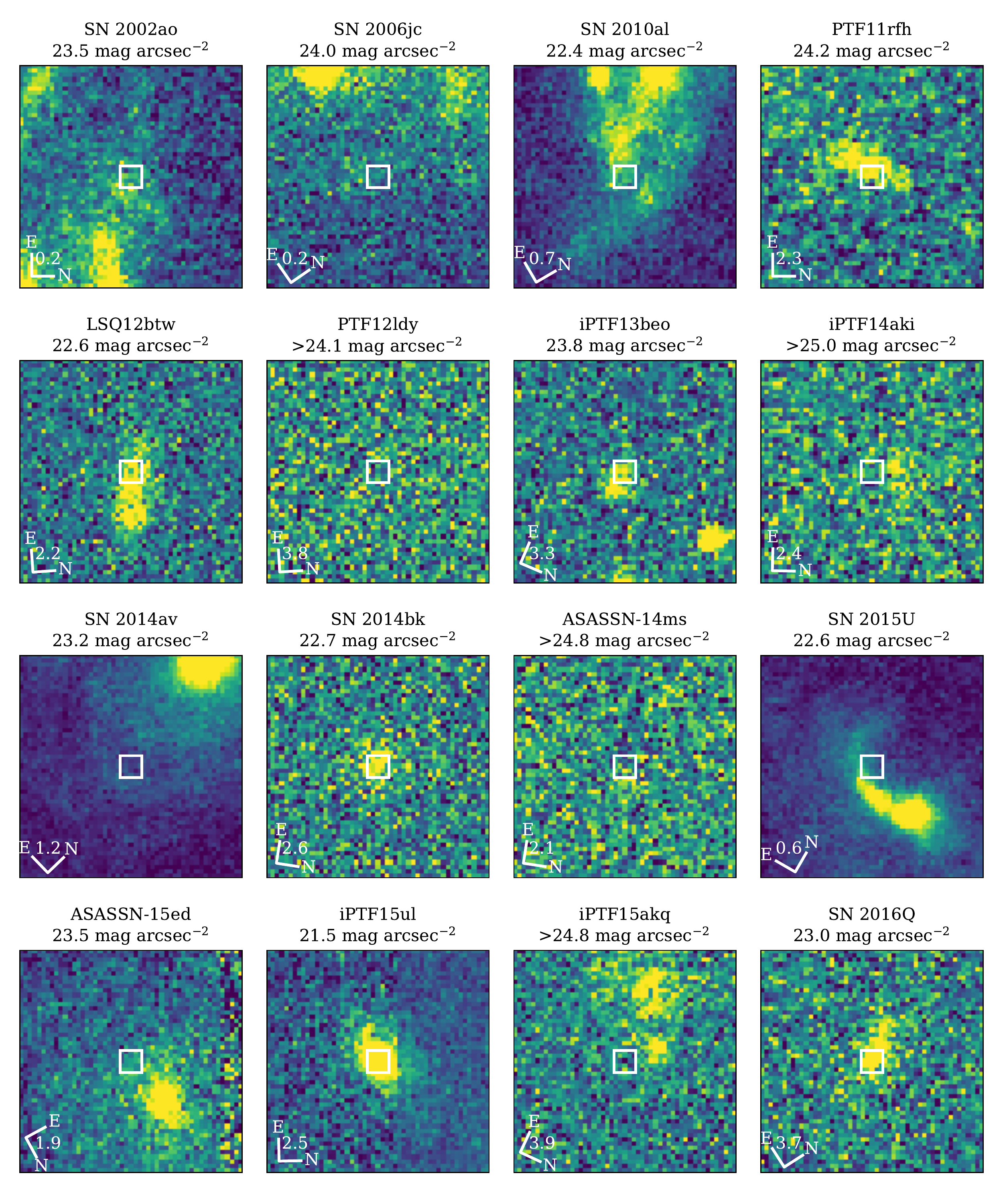}
    \caption{$u'$-band images of SN~Ibn host galaxies from SDSS. We measured the surface brightnesses within the $5 \times 5$ pixel ($1\farcs98 \times 1\farcs98$) white square centered on the SN location. The compass rose at the bottom left of each panel shows the orientation and physical size in kpc of the aperture. PS1-12sk is omitted because the limit from \textit{HST} is much deeper. We did not measure any flux for four of the SNe~Ibn, but the $3\sigma$ nondetection limits are not very constraining (see Figure~\ref{fig:limits}).}
    \label{fig:hosts}
\end{figure*}

We followed a methodology similar to \cite{Kelly2012} to measure $u'$-band surface brightnesses at the locations of the 17 SNe~Ibn whose host galaxies were observed by SDSS \citepalias{SDSSCollaboration2017}.\footnote{SN~2005la was an additional SN~Ibn (actually a transitional Type~Ibn/IIn) in an SDSS field, but that image is masked due to a nearby 5~mag star.} Because we include all classified SNe~Ibn in a predefined area of the sky, this should be an unbiased sample. Indeed, we do not observe any correlation between our measured surface brightnesses and redshift. We downloaded the background-subtracted $u'$-band images from SDSS, sometimes multiple frames per field, using Astroquery \citep{astroquery} and calculated variance images according to the SDSS \href{https://data.sdss.org/datamodel/files/BOSS_PHOTOOBJ/frames/RERUN/RUN/CAMCOL/frame.html}{frame data model}. We stacked the images of each host, weighting by inverse variance, and performed aperture photometry on the stacked image and the total variance image using a $5 \times 5$ pixel ($1\farcs98 \times 1\farcs98$) square aperture centered on the location of the SN. Although the physical size of this aperture varies by more than an order of magnitude within our sample, we chose it as a compromise between increasing our statistical uncertainties by including too few pixels and increasing our systematic uncertainties by including significant surface brightness gradients, where the relevant scale for the latter is set by the typical SDSS seeing of $1\arcsec$. In comparison, \citealt{Kelly2012} used a 20 pixel aperture. $2\arcsec$ also approximately matches the 2~kpc aperture we use for the \textit{HST} images of PS1-12sk in Section~\ref{sec:obs}. The results are listed in Table~\ref{tab:hosts} and shown in Figure~\ref{fig:hosts}.

Equations~(\ref{eq:sfr})--(\ref{eq:sfrd}) are not strictly applicable to these measurements, because the SDSS $u'$ filter does not overlap with the 280~nm filter used to derive them. However, $u'$-band luminosity is still dominated by young stars, but with significant contamination from older populations \citep{Hopkins2003}. \cite{Hopkins2003} calibrated star formation rate to $u'$-band luminosity, but because their relationship is nonlinear, we cannot rewrite it in terms of star formation rate density and surface brightness. \cite{Cram1998} produced a linear calibration between $U$-band luminosity and the formation rate of $>$5~M$_\sun$ stars only. To convert this to a full star formation rate, we multiply by the ratio of mass from stars $>$0.5~M$_\sun$ to mass from stars $>$5~M$_\sun$ according to the \cite{Salpeter1955} initial mass function, which we find to be 22. This ratio is very sensitive to the 0.5~M$_\sun$ lower limit but results in a relationship very close to Equation~(\ref{eq:sfr}). Rather than trust this modified calibration, we take this to mean that Equation~(\ref{eq:sfr}) can be used with $u'$-band measurements and apply it to our results in Table~\ref{tab:hosts}. Because all of the hosts are at redshifts $z < 0.11$, we again neglect $K$-corrections.
 
\begin{figure*}[t]
    \centering
    \includegraphics{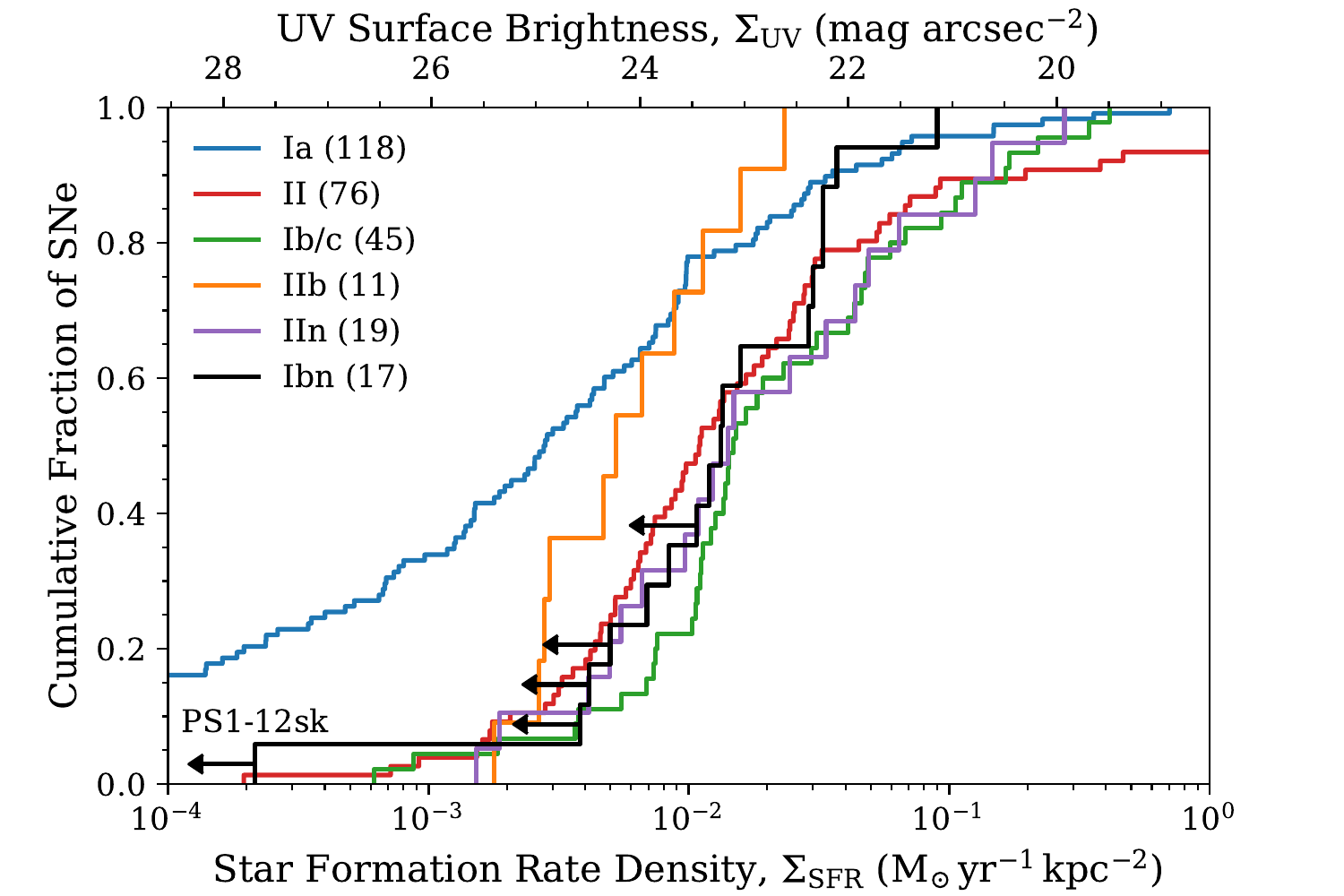}
    \caption{Comparing the star formation rate density limit at the location of PS1-12sk to other SNe~Ibn and the samples of \cite{Galbany2018}, with sample sizes in parentheses. We omit the transitional Type~Ibn/IIn SN~2005la from the sample of SNe~IIb. The upper axis shows UV surface brightness corrected for redshift and Milky Way extinction. Left-pointing arrows indicate $3\sigma$ upper limits. Our limit for PS1-12sk rules out all previous core-collapse SN hosts in this sample at $>\!2.7\sigma$.}
    \label{fig:limits}
\end{figure*}

Twelve events have $u'$-band surface brightnesses, and therefore star formation rate densities, well within the normal range for core-collapse SNe (see Figure~\ref{fig:limits}). We do not see any SNe~Ibn in very bright regions, but this could be an observational bias: rare and fast-evolving transients are harder to find and classify near galaxy centers. We do not detect flux at the locations of the remaining five SNe~Ibn, but the limits are not very constraining. Apart from PS1-12sk, these occurred near very faint SDSS galaxies. In fact, \cite{Vallely2018} measured a star formation rate for the host galaxy of ASASSN-14ms of $\sim$0.02~M$_\odot$~yr$^{-1}$.
 
Figure~\ref{fig:limits} compares our results for SNe~Ibn to star formation rate densities at the locations of other SNe from the sample of \cite{Galbany2018}, who used H$\alpha$ luminosity from IFU spectra as a diagnostic. Note that \citet[p.\ 3]{Galbany2018} ``aimed to resolve a bias identified in \cite{Galbany2014} due to the absence of low-mass galaxies in the CALIFA sample'' and that all their measurements use a 1~kpc aperture, which is close to the average aperture in our SDSS sample (see Table~\ref{tab:hosts}) and close to the 2~kpc aperture we use for the \textit{HST} images PS1-12sk. By comparing our star formation rate densities with theirs, we are implicitly assuming that our undetected hosts are larger than the aperture size and/or of similar physical size to the comparison samples, although an extremely small host would be interesting in itself. All of their core-collapse SN hosts are detected in H$\alpha$; SN~Ia hosts that are not detected are plotted at $\Sigma_\mathrm{SFR} = 0$. The faintest core-collapse host is that of the Type~II SN~1948B, with a star formation rate density about equal to our $3\sigma$ limit for the host of PS1-12sk.\footnote{Following advice from L.~Galbany (2018, private communication), we correct the erroneous value for SN~1948B to $\log \Sigma_\mathrm{SFR} = -3.7088$.} However, this is just the extreme of a well-sampled distribution of SNe~II, whereas PS1-12sk is clearly an outlier among the smaller sample of SNe~Ibn. If we exclude PS1-12sk, the there would be no evidence that the distribution of SNe~Ibn was any different than the distributions for other types of core-collapse SNe.

For completeness, we remark on the host galaxies of the four SNe~Ibn that have been discovered in the past two years, as they have not been mentioned elsewhere.
SN~2017ecp \citep{2017TNSCR.579....1H,2017TNSTR.559....1K} occurred $0\farcs7$ from the center of the spiral galaxy PGC~727891 \citep{hyperleda}.
SN~2017iwp \citep{2017TNSTR1399....1D,2017TNSCR1427....1S} occurred $10\farcs8$ from the center of ESO~162-G7 \citep{1998yCat.7034....0L}, which is classified as an S0 galaxy but is small enough ($1\arcmin \times 0\farcm1$) that spiral structure might be unresolved.
SN~2017jfv \citep{2017TNSTR1485....1D,2018TNSCR..19....1B} occurred $6\farcs4$ from the center and matches the redshift of the emission-line galaxy ESO~444-IG49 NED01, which is part of a galaxy pair \citep{1998yCat.7034....0L}.
SN~2018fmt \citep{2018TNSTR1262....1F,2018TNSCR1339....1G} occurred $1\farcs3$ from the center and matches the redshift of the UV-bright galaxy GALEXASC~J012100.76-135145.0 \citep{Maddox1990}, which is part of the galaxy cluster MXCX~J0120.9-1351 \citep{MCXC}.

\section{What Produced PS1-12sk?} \label{sec:interp}
Our limits at the location of PS1-12sk rule out all but one of the core-collapse SN hosts in the sample of \cite{Galbany2018} at $>\!5\sigma$ significance; we rule out the host with the lowest star formation rate at $2.7\sigma$ (Figure~\ref{fig:limits}). There are therefore three remaining possibilities for the origin of PS1-12sk.

\subsection{Option 1: Ejected or Stripped from Ultra-compact Dwarf}
The progenitor of PS1-12sk could have originated in the ultra-compact dwarf, in which case it would have had to travel more than 2.7~kpc to the SN location before exploding. Applying Equation~(\ref{eq:sfr}) to the luminosity that we measure in Section~\ref{sec:obs}, we calculate a star formation rate of $5.1 \times 10^{-3}\,\mathrm{M_\odot\,yr^{-1}}$ for the dwarf. \cite{Botticella2012} gave the fraction of stars per unit mass that produce core-collapse supernovae as $K_\mathrm{CC} \sim 0.01\mathrm{\,M_\odot^{-1}}$, meaning that we would expect a massive star from this galaxy to explode once every $\approx$20~kyr. However, at over 22.5 Kron radii, this would be by far the largest normalized offset for a core-collapse SN: the largest in the sample of \cite{Kelly2012} is 8.77 half-light radii, and all the SNe~Ibn examined by \cite{Pastorello2015a} occurred within the 25~mag~arcsec$^{-2}$ isophotal radius. The only SN~Ibn with a direct progenitor detection (albeit in outburst) was SN~2006jc; \cite{Tominaga2008} estimated that this SN came from a $M_\mathrm{ZAMS} = 40\,\mathrm{M}_\odot$ star, which would have a lifetime of $\approx$5~Myr. To move $>$2.7~kpc during that time, a star this massive would need a velocity $>$530~km~s$^{-1}$.

Encounters with supermassive black holes can accelerate stars up to velocities of several hundred km~s$^{-1}$ \citep{Hills1988}.  This is unusually fast for stars in the Milky Way \citep{Brown2007}, but well above the escape velocity for an ultra-compact dwarf. Recent observations of the velocity dispersion profiles of two ultra-compact dwarfs do suggest that they have central black holes of the required mass \citep{Ahn2017}. Alternatively, the BCG or another cluster member could have stripped the star from the ultra-compact dwarf during a tidal interaction, in which case the star's velocity would be consistent with the velocity dispersion of the cluster \citep{Moore1996,Sommer-Larsen2005,DeMaio2018}. \cite{Bohringer2000} gave the cluster X-ray luminosity as $L_\mathrm{X} = 0.16 \times 10^{44}\mathrm{\,erg\,s^{-1}}$, implying a velocity dispersion around 500~km~s$^{-1}$ \citep{Zhang2011}. Its relatively low X-ray luminosity makes it more likely to have tidal interactions and mergers \citep{Zabludoff1998}. In fact, this cluster's BCG is likely the result of such mergers \citep{Eigenthaler2009}.

Although both of these processes are physically possible, we must keep in mind that this would be the first core-collapse SN discovered in the intracluster medium. For it to be of such a rare class (Type~Ibn) suggests some connection between the progenitor system and its escape mechanism from the ultra-compact dwarf. SNe~Ia, on the other hand, have been found in the intracluster medium before \citep{Sand2011}.

\subsection{Option 2: Luck}
We may have observed a very statistically improbable event. There could be a very low level of star formation at the SN location, beneath our very deep limits, and we happened to see one of the few massive stars that formed there explode. Even if there was more vigorous star formation that suddenly shut off sometime in the past $\approx$5~Myr, we would still expect to see UV emission today; the mean stellar age that contributes to this emission is 10~Myr \citep{Kennicutt2012}. This star formation could be associated with the BCG, the intracluster medium, or a gas flow into the cluster core. Previously, only a single core-collapse SN has been observed in an elliptical cluster galaxy, the SN II Abell399\_11\_19\_0 \citep{Graham2012}.

As a rough estimate of the likelihood of this observation, we can multiply our limit on the star-formation rate by $K_\mathrm{CC} \sim 0.01\mathrm{\,M_\odot^{-1}}$ \citep{Botticella2012} to get a rate of a few per Myr kpc$^2$. Because SNe~Ibn make up only about 1\% of core-collapse SNe \citep{Pastorello2008a}, it is especially unlikely that the second core-collapse SN that we observed in such an environment was of a very rare type. We would expect this a few times per 100 Myr kpc$^2$. If we assume the progenitor could have moved at most a few kpc during its lifetime (see previous section) and integrate over this region, we get a rate of SNe~Ibn of $\lesssim\!1\,\mathrm{Myr}^{-1}$.

\subsection{Option 3: Low-mass Progenitor}
The last remaining option is that PS1-12sk did not come from a massive star. In other words, we might not detect any star formation at the SN location because it ceased billions of years ago. The exploding star could then either be a white dwarf in a helium envelope or a nondegenerate helium star. For example, \cite{Sanders2013} suggested that explosions of stars like AM Canum Venaticorum and R Coronae Borealis could reproduce some of the observed features of SNe~Ibn, but neither is completely satisfactory. Likewise, calcium-rich transients \citep{Perets2010} share many features with PS1-12sk---spectra showing helium but not hydrogen, rapidly fading light curves, some host galaxies with no star formation, and large host offsets---but are much fainter and, crucially, do not show narrow spectral lines indicating circumstellar interaction.

One of the strongest pieces of evidence that SNe~Ibn come from massive stars is the pre-explosion eruption observed two years before SN~2006jc, which peaked at $M_R \approx -14$~mag and lasted for $\approx$10~days \citep{Pastorello2007}. We are not aware of any way to produce this phenomenon with a low-mass, hydrogen-poor star. Helium novae are orders of magnitude too faint \citep{Kato2003}. Theoretical .Ia SNe roughly match the required luminosity and time scale, but we would not expect their progenitors to explode two years later as SNe~Ia \citep{Bildsten2007}. The fast-declining light curves of SNe~Ibn also indicate that very little $^{56}$Ni is produced in the explosion \citep{Foley2007,Pastorello2008a}, which may challenge white dwarf explosion models.

One way of avoiding these issues is to allow PS1-12sk and SN~2006jc to come from different types of stars. SNe~Ibn would then be a mixed class, with some low- and some high-mass progenitors. This could potentially explain some of the spectral diversity we see among SNe~Ibn \citep{Hosseinzadeh2017c}. The five spectra that we have of PS1-12sk look almost identical to spectra of SN~2006jc,\footnote{SN~2006jc was discovered after peak, so the phases of these spectra are not well constrained.} but perhaps any explosion into helium-rich CSM would look the same. This situation would be analogous to the discovery that some events initially classified as SNe~IIn are actually the explosions of white dwarfs with hydrogen-rich CSM \citep[now called SNe~Ia-CSM;][]{Dilday2012,Silverman2013}. In fact, this may be the reason that events classified as SNe~IIn are less correlated with star formation than other classes of core-collapse SNe \citep{Anderson2012,Habergham2014}, despite some having luminous progenitor detections.

\section{Summary and Conclusions}
We have presented a very deep limit on the star formation rate at the site of the SN~Ibn PS1-12sk, nominally ruling out a massive star origin. We present three alternative proposals for how this SN could have occurred in such a unique environment.
\begin{enumerate}
    \item The progenitor of PS1-12sk was a massive star ejected or tidally stripped from a nearby ultra-compact dwarf galaxy.
    \item The progenitor of PS1-12sk was a rare massive star in a region with a very low star formation rate.
    \item The progenitor of PS1-12sk was a low-mass star.
\end{enumerate}

Our result only deepens the mystery surrounding PS1-12sk, in that none of these possibilities is fully satisfactory in explaining the origins of SNe~Ibn. Given that there is not much more data to collect on this particular event, the resolution will likely come from a statistical study of SN~Ibn host galaxies. Although this is currently the only SN~Ibn discovered in a non-star-forming environment, we still suffer from a small sample size of only tens of events. Current and future large-scale surveys will be much better suited to find these rare, fast-evolving transients, allowing us to better understand the statistical properties of their host environments.

\acknowledgements
We thank Llu\'is Galbany for advice regarding his core-collapse host galaxy measurements, Charlotte Mason for advice on galaxy photometry, and Lars Bildsten, Jim Fuller, and Joe Anderson for useful discussions about possible SN~Ibn progenitors. G.H.\ and D.H.\ were partially supported by NASA through grant No.\ HST-GO-15236.001-A from the Space Telescope Science Institute. A.I.Z.\ acknowledges support from NSF grant AST-1715609. K.D.F.\ is supported by program No.\ HST-HF2-51391.001-A, provided by NASA through a grant from the Space Telescope Science Institute. G.H.\ thanks the LSSTC Data Science Fellowship Program, which is funded by LSSTC, NSF Cybertraining Grant \#1829740, the Brinson Foundation, and the Moore Foundation; his participation in the program has benefited this work.

\appendix
\section{Extinction}\label{sec:extinct}
\citet[p.\ 4]{Sanders2013} claimed that ``the SED of [PS1-12sk] is not consistent with significant reddening given reasonable assumptions about the photospheric temperature'' and neglect extinction throughout their analysis. However, as an additional check, we calculate an upper limit on the host galaxy extinction based on the nondetection of sodium absorption in a spectrum of the SN. The highest-resolution spectrum available was taken 8~days after maximum light using MMT's Blue Channel spectrograph with the 832 line mm$^{-1}$ grating \citep[$\Delta x = 0.072\mathrm{~nm~pixel}^{-1}$; resolution = 0.204~nm;][]{Schmidt1989}. It is not plotted by \cite{Sanders2013}, but it is available for download from WISEeREP \citep{Yaron2012}.

We follow a procedure similar to \cite{Leonard2001}. The sodium lines used to measure extinction (589.0 and 589.6~nm) are very near one of the spectrum's strongest emission features (helium 587.6~nm), but because the sodium lines would be much narrower than the emission feature, this does not interfere with our analysis. We fit a second-order \cite{Savitzky1964} filter with a width of 99 pixels to produce a smoothed spectrum and normalize the original data by the smoothed spectrum. We then calculate the standard deviation of the normalized spectrum ($\Delta I$) within 5 resolution elements of the sodium lines. The measurement does not change significantly if we use between 3 and 7 resolution elements. Applying Eq.~4 of \cite{Leonard2001}, we get a $3\sigma$ limit on the equivalent width of sodium of $W_\lambda < 3 \Delta I \sqrt{W_\mathrm{line} \Delta x} = 0.0061$~nm, where we have assumed the width of the line is $W_\mathrm{line} = 1$~nm. This corresponds to an extinction coefficient of $A_\mathrm{F300X} < 0.088$~mag \citep{Cardelli1989,Barbon1990}. At most, extinction could weaken our limit by about 8\%.

\section{Background Subtraction}\label{sec:bkg}
Because we are attempting to detect a low surface brightness feature in a galaxy cluster environment, careful measurement and subtraction of the background is crucial. Unfortunately, the background flux varies between the four detector quadrants (because each has its own amplifier) as well as across each quadrant. Quadrants A and B are not used in our analysis except for in refining the astrometric solution of the images before stacking. The BCG falls in quadrant C, and the SN location lies very near the quadrant boundary, such that our measurement aperture is spread across quadrants C and D. We want to remove any instrumental artifacts while preserving any light intrinsic to the galaxy cluster.

We model the instrumental signature by fitting a smoothly varying 2D background to the full image using \texttt{sep.Background} \citep{Barbary2016} with a box size of 64 pixels. This is approximately the same size as the BCG, so we mask an $800 \times 800$~pixel region centered on the BCG before doing the fit. From this model, we observe that the background varies smoothly by 5--10 electrons from the outer edges of the chips (brighter) toward the chip gap (fainter), but it does not vary significantly in the direction parallel to the chip gap, except for a small shift at the quadrant boundaries. We do not see any sign of intracluster light (an excess centered on the BCG) in our background model, so we proceed to subtract this model from the F300X image before doing any measurements.

The only number that quantitatively affects our analysis is the average value of the background inside our aperture. This determines whether or not the flux measurement surpasses our 5$\sigma$ threshold for a detection. We explored using several other values for this quantity: the median in quadrant C, the median in quadrant D, and the median in the region shown in Figure~\ref{fig:images}, all calculated using both the original image and the background model. None of these seven options yields a detection with more than $1.2\sigma$ significance, which, if real, would still be the lowest star formation rate density for a core-collapse host. As we are confident that there is no excess, we calculate an upper limit based on the Poisson uncertainty on the sum within the aperture, which is not affected by the background model.

\facilities{ADS, HST (WFC3), NED.}
\defcitealias{AstropyCollaboration2018}{Astropy Collaboration 2018}
\software{Astropy \citepalias{AstropyCollaboration2018}, Astro-SCRAPPY \citep{McCully2014}, Astroquery \citep{astroquery}, DrizzlePac \citep{Gonzaga2012}, SciPy \citep{scipy}, SEP \citep{Barbary2016}, SNHST \citep{McCully2018}.}

\end{document}

%% file: affil.tex
\newcommand{\LCO}{\affiliation{Las Cumbres Observatory, 6740 Cortona Drive, Suite 102, Goleta, CA 93117-5575, USA}}
\newcommand{\UCSB}{\affiliation{Department of Physics, University of California, Santa Barbara, CA 93106-9530, USA}}

\newcommand{\CfA}{\affiliation{Center for Astrophysics \textbar{} Harvard \& Smithsonian, 60 Garden Street, Cambridge, MA 02138-1516, USA}}
\newcommand{\UA}{\affiliation{Department of Astronomy/Steward Observatory, 933 North Cherry Avenue, Tucson, AZ 85721-0065, USA}}

\newcommand{\DSFP}{\altaffiliation{LSSTC Data Science Fellow}}

\newcommand{\Carnegie}{\affiliation{Observatories of the Carnegie Institute for Science, 813 Santa Barbara Street, Pasadena, CA 91101-1232, USA}}
\newcommand{\TAU}{\affiliation{School of Physics and Astronomy, Tel Aviv University, Tel Aviv 69978, Israel}}